# New thermodynamic regularity for cesium over the whole liquid range

M.H. Ghatee* and M. Bahadori

(Department of Chemistry, Shiraz University, Shiraz, 71454, Iran)

Email: ghatee@sun01.susc.ac.ir

Fax: +98 711 2286008

*Abstarct*

*In this paper we derive an equation of state for liquid cesium based on a suggested potential function in accord to the characteristics large attraction and soft repulsion at the asymptotes of interaction potentials. By considering the interaction of nearest adjacent atoms in dense fluid, the equation of state predicts that the isotherm $(Z-1)V^2$ is linear function of $(1/\rho)$, where $Z$ is the compression factor, $V$ is the molar volume, and $\rho$ is the molar density. The linear parameters are identified as interaction coefficients related to attraction and repulsion, and are used to evaluate the molecular parameters with interesting implications. The isotherm is intended to resolve the particular thermodynamic properties of alkali metals, which have been known for their unusual change of the nature of intermolecular force as the characteristic metal-nonmetal transition range is approached. When applied to liquid cesium, the isotherms persist linear over the whole liquid range including the metal non-metals transition range and at the critical temperature perfectly. The isotherm is equivalent to a virial (like) EOS for which the linear parameters of the isotherm form the corresponding second and third virial coefficients. The new potential function turns out to be an effective potential that includes not only a pair interaction but also many-body interactions and therefore it is not always comparable with pair potential.*

# 1. Introduction

The electronic structure of liquid metals at low temperatures can be well approximated by the structure of the solid metals, and the interaction between atoms and thus the thermodynamic properties of metals are obtainable by the same cohesion mechanism of the solid metals. The structure of a solid metal is usually considered as a collection of ions fixed at the solid matrix imposed by the solid structure, and the valance electrons delocalized over the whole lattice having little correlations with core electrons. At ordinary liquid density near triple point, the effective interaction potential of alkali metals can be obtained by pseudopotential perturbation theory based on nearly free electron model. This effective procedure considers the nuclei completely shielded by core electrons and partially by delocalized valence electrons. Near critical point, however, the cohesion mechanism of these metals will be suppressed by a (partial) localization of electrons and the metallic character is gradually changed to a nonmetallic kind. In the gas phase, on the other hand, the metal vapor is composed of mainly neutral atoms plus ions and molecules comprising clusters of different size. The interaction between these atom and clusters can be modeled to a great deal by Lennard-Jones potential function. But, near the critical point especially in the metal-nonmetal transition region, there is no reliable theoretical method to derive effective potential function for liquids accurately. Indeed, the effective pair potential function for liquid at high densities near the triple point, obtained from nearly free electron model, can reproduce the thermodynamic properties quite well, though, it gives less satisfactory results in the expanded liquid metals at low densities (near critical point). Under such circumstance the accuracy of thermodynamic properties of liquid metals, predicted by theoretical studies, depends on the accuracy of pair potential function describing the interatomic interaction of these metals.

The electronic structure effectively determines the interatomic interaction and material structure as well, and therefore a unified understanding of the structure, the thermodynamics, and electronic properties of metal fluids, emanates a considerable scientific challenge. A great deal of effort has been devoted to alkali metals to study the marked change in the nature of forces, as liquid metals are expanded.[1,2,3] Close to critical temperature, the change in the nature of forces in liquid alkali is characterized by a sharp decrease in electrical conductivity. Structure factors and pair correlation functions based neutron scattering data show that the coordination numbers of these metals drop linearly with temperature,[2] and accordingly, the nearest adjacent distance increases steadily. These marked properties make the theoretical studies ambiguous and a meaningful application of pair potential for dense fluids need a pronounced justification. There has been an open field to describe interaction potentials of metals by conventional models and considering multipolar interactions, which are expressed as multipole expansion series to describe the attractive branch of the interaction potential as accurate as possible.[4]

In a series of works, the Linear Regularity Isotherm (LIR),[5] which originally devised for normal fluids, has been applied to the alkali metals.[6,7] LIR is based on the cell theory and considers only the nearest adjacent interaction. In the final form it formulates isotherms of $(Z-1)V^2$, which are linear function of $\rho^2$, where $Z = P/\rho RT$ is the compression factor, $V$ is the molar volume, $\rho$ is the molar density, $P$ is the pressure, and $RT$ has its usual meaning. LIR is applied well to the most Lennard-Jones fluids which their interaction potential can be modeled rather accurately by the (12-6) powers of inverse intermolecular distance. This is because the nature of forces of Lennard-Jones fluids are described well according to a definite dispersive interaction mechanism. However, in the application of LIR to alkali metals liquids, isotherms deviate from the linear behavior as the critical temperature is approached. Of course, the application of LIR to alkali metals, as explored for the first time for liquid cesium,[6] suffers from metal-nonmetal transition as a consequence of the diversion of interatomic forces in the critical temperature region. A conclusive remark was that the assumption of the nearest adjacent interaction in the derivation of the LIR is break down.[6,7] These observations have been attributed basically to onset of the localization and partial correlation of valence electrons as the critical temperature is approached, which lead to formation of metal-atoms together with poly-atom clusters.[7] Based on this objection, liquid alkali metals has been treated statistical mechanically as an ensemble of mixture of metal-atoms and poly-atom clusters,[7] and shown that LIR leads to isotherm $(Z-1)V^2$ which is a quadratic function in $\rho^2$. Although the composite specification of the mixture was not specified but, by this modeling, the deviations of isotherms from a linear behavior (in the transition region) have been attributed to the fractionation of the atomic liquid metal system into molecular clusters.

Following the explanation of the cause of transition, the linear exp-6 isotherm[8,9] has derived by applying the exp-6 interaction potential function to model the long-range and the short-range interactions in the liquid alkali metals more realistically. Accordingly, isotherms of $(Z-1)V^2$ has found to be linear function of $\rho^{-7/3}\exp[\alpha(1-K\rho^{-1/3}/r_m)]$, where $K$ is a constant characteristic of the unit cell structure of the metal, $r_m$ is the equilibrium interatomic distance, and $\alpha$ is a constant that adjusts asymptotes of both repulsion and attraction branches of the exp-6 potential function. By application of exp-6 isotherm to liquid cesium, the linear feature persists well beyond the onset of the metal-nonmetal transition region at which the transition is admitted. [The metal-nonmetal transition has been identified at 1450 K by LIR, and consistently at $1.21\,\mathrm{gr/cm^3}$ by electrical conductivity measurement].[1-3] The success of the exp-6 isotherm has been mainly attributed to its well-estimation of the attraction potential energy,[8] which is increased as a result of increase in the number of polarized metal-atoms (forming clusters) as the liquid metal is expanded. In fact, the large polarizablity of a metal-atom, in the low-density liquid, is due to the increase in correlation of the single valence electron with core electrons of the respective ion.

The equilibrium interatomic distance $r_m$ appears as a parameter in exp-6 isotherm. This plus the fact that $r_m$ is a function of temperature limits the applicability of the exp-6 isotherm for practical and theoretical purposes.

In this paper, we explore further the idea that the cause of deviation of (LIR) isotherms in expanded liquid alkali metals in the transition region is due to the under-estimation of extra attractive potential due to development of polarized metal-atoms in addition to ploy-atom clusters. According to neutron scattering data and application of an inversion method for interaction potential energy in liquid cesium, we present a two-parameter pair (type) potential function. It is furnished to account suitably for the repulsion characteristics of alkali metals with soft electron clouds and the attraction characteristics of systems with appreciable coulombic long-range attraction. The equation of state, derived accordingly, is presented as a linear isotherm, which on rearrangement would be identical to a virial equation of state up to third virial, where the linear parameters of the isotherm are the interaction (virial) coefficients. The isotherm by this modeling is applied to liquid cesium for which PVT data are available over the whole liquid range. The temperature dependence of parameters of the pair potential parameters are examined and its physical significant in relation to liquid structure will be presented.

## 2. Potential function

Lennard-Jones pair potential function may be represented in general form as

$$u(r) = A\varepsilon\left[\left(\frac{\sigma}{r}\right)^m - \left(\frac{\sigma}{r}\right)^n\right] \tag{1}$$

where $A$ is a constant, $\varepsilon$ is the potential well depth, and $\sigma$ is the diameter hard molecule. The potential function of Eq. (1) obeys boundary conditions such that

$$r_m = \left(\frac{m}{n}\right)^{1/(m-n)}\sigma, \qquad A = \left(\frac{m}{m-n}\right)\left(\frac{m}{n}\right)^{n/(m-n)}. \tag{2}$$

For a large number of atomic and molecular fluids, Lennard-Jones (12-6) potential function accounts for the pairwise interaction approximation between the fluid molecules undergoing dispersive interaction as the major interaction. The interest in Lennard-Jones is also due to its simple form, which facilitates simple integration and differentiation desired for derivation of analytical form of thermodynamic functions. However, in some typical dense fluid Lennard-Jones loss its application because it introduces unrealistic hard repulsion. Many fluids also interact with an appreciable coulombic feature so much the long-range dispersive interaction potential is under-estimated if treated by only a simple sixth power of inverse of interatomic distance.

In spite of the progress in understanding of the electronic properties of expanded alkali metals, essentially not very much is known on the strong thermodynamic state dependence of their effective interatomic interaction especially the change in these interaction as the metal-nonmetal transition is approached.[10] The recent neutron scattering experiments for cesium as a function of temperature and pressure, however, directed to investigate the more detailed density dependence of the effective interaction potentials of expanded alkali metals.[10] This requests a theory that further relates the structure factor to the interaction potential.[10,11] Results of the undertaken neutron scattering as a function of pressure have shown a change in the position of the first peak of the pair correlation function, denoting a slight change in the mean interatomic distance. The analysis of data in general has shown that at high densities an alkali metal atom interacts with a repulsive soft-core at small interatomic distances and with a weak attraction at intermediate interatomic distance. At larger atomic separation the potential function oscillates and takes $(1/r^3)$ asymptotic form. On the other hand, by the integral equation theory, the effective pair potential of liquid cesium for a wide range of density has been derived from the experimental structure factors using the inverse method.[11] The derived pair potential function at low temperatures oscillates at large interatomic distances but, as temperature is increased, the oscillations are dumped out.[11] Also as the temperature is increased the long-range interaction would be characterized by a more attraction. The repulsive side of this potential has been analyzed in terms of an inverse power-law $(a/r^m)$, where $a$ is a constant.[11,12] For cesium $m = 7.7, 6.8$ and $5.6$, at $773\,\text{K}, 1373\,\text{K},$ and $1673\,\text{K}$, respectively. For rubidium the corresponding m values are larger indicating the larger the size of an alkali atom the softer is the electronic cloud. It is well known that the interionic dipole-dipole interaction has an effect of softening the pair potential.[13,14] Since the ionic polarizabilitiy of cesium is larger than other alkali metals, the softening is also expected to be larger for cesium than other alkali metals.

Form the forgoing analysis of the experimental data and from the fact that the range of liquid metal densities correspond to the interatomic distances mainly located around the potential minimum and likely is extended to larger atomic distances than $r_m$, we propose that the pair potential

$$u(r) = A\varepsilon \left[ \left(\frac{\sigma}{r}\right)^6 - \left(\frac{\sigma}{r}\right)^3 \right] \qquad (3)$$

would account for the interatomic interaction to model and thus to predict the thermodynamic properties of liquid cesium significantly accurate. In Eq. (3) $A = 4$ $\sigma = 2^{1/3} r_m$ and $r_m$ is the position of potential minimum. Of course, we propose the semiempirical potential Eq. (2) to direct the idea of presenting a simple potential function that could accounts, to a good approximation, for the asymptotic behavior of both attraction and repulsion potential over the whole liquid range corresponding to the intermediate interatomic distances. This is of major concern in this work to investigation the detail of a potential function suitable for liquid alkali metals (in particular liquid cesium) in the whole liquid range, particularly in the range that a liquid metal turns to the corresponding nonmetal.

## 3. Linear Isotherm

From the discussions in the preceding section, we believed that the potential function Eq. (3) could suitably describes the interaction potential between pair of cesium metal atoms in dense liquid. Thus, we apply it to evaluate the interaction potential energy of atoms in the liquid assuming that liquid obey the following model. Close to freezing temperature, the liquid alkali metals have a body center cubic structure, closely similar to their corresponding solids structure. Therefore, any cesium atom has 8 nearest adjacent atoms, and by assuming the pairwise additivity of the interaction potential the molecular parameters in Eq. (1) or (3) would be different from the molecular parameters of an isolated pair. In this model, we assume the complete pairwise additivity of the interaction potential between any atom at the center and its nearest adjacent atoms. Therefore, the total potential energy U of N-atom liquid is calculated as

$$U(\mathbf{r}_1,...,\mathbf{r}_N) = \sum_{i>j=1}^{N} u(\mathbf{r}_i,\mathbf{r}_j) \quad (4)$$

where the pair potential $u(\mathbf{r}_i,\mathbf{r}_j)$ is often assumed to depend only on the distance $\mathbf{r}_{ij} = |(\mathbf{r}_i - \mathbf{r}_j)|$ between the $i$th and $j$th pair of molecules located at positions $\mathbf{r}_i$ and $\mathbf{r}_j$, respectively. Furthermore, we assume that any atom interact with its nearest adjacent atoms in pairs, like a pair of atoms 1 and 2 with pair potential $u(\mathbf{r}_{12})$ and thus, we calculate the potential energy as

$$U(\mathbf{r}_1,...,\mathbf{r}_N) = \frac{N}{2} u(\mathbf{r}_{12}) \quad (5)$$

where the factor $1/2$ is used to follow the restriction $i > j$ – excluding all identical interatomic interactions in the summation. Thus in our model $\varepsilon$ is binding energy of atoms 1 and 2 in the ensemble of N-2 other atoms, knowing that the most effective interaction are due to the nearest adjacent atoms. In this way, insertion of Eq. (3) into Eq. (5) for liquid metals may yield $\varepsilon$ as binding energy of an atom effectively interacting with atoms on its nearest adjacent shell (and next shells as well). This concept of biding energy is well justified when the experimental PVT data is used to evaluate the molecular potential parameters.

If Eq. (5) can be evaluated definitely, e.g., using an accurate $u(\mathbf{r}_{12})$, the thermodynamic equation of state can be solved to estimate accurate mechanical pressure P:

$$P = T\left(\frac{\partial P}{\partial T}\right)_V - P_{int} \tag{6}$$

Now, the internal pressure $P_{int}$ is evaluated, using Eqs. (3) and (5)

$$P_{int} = \left(\frac{\partial (U+E_K)/N}{\partial V}\right)_T \approx -C_1\rho^3 + B_1\rho^2 \tag{7}$$

where $E_K$ is the kinetic energy, and the constants $C_1$ and $B_1$ are in terms of molecular parameters. Accordingly,

$$C_1 = \frac{4N\varepsilon\sigma^6}{K^6}, \qquad B_1 = \frac{2N\varepsilon\sigma^3}{K^3}, \tag{8}$$

where $K\left(=\left(3\sqrt{3}/4N\right)^{1/3}\right)$ is raised by considering the structure of liquid as body-center-cubic.

Notice that we don't need to know the form of kinetic energy $E_K$ because it is independent of volume. Then the equation of state, subjected to the boundary conditions of Eq. (3), and as explicit function of temperature, is presented by the isotherm,

$$(Z-1)V^2 = C + B\left(\frac{1}{\rho}\right) \tag{9}$$

where Z is the compression factor. Here $C$ and $B$ are temperature dependent constants of the isotherm and are derived as

$$C = \frac{1}{\rho^2}\left[\frac{1}{\rho R}\left(\frac{\partial P}{\partial T}\right)_\rho - 1\right] + \frac{C_1}{RT} = C_2 + \frac{C_1}{RT}, \qquad B = \frac{-B_1}{RT}. \tag{10}$$

In Eq. (9) molar quantities are applied where applicable. The Eq. (9) is derived similar to our previous work, but it is different from the original LIR in that we have employed the potential function Eq. (3) and have considered $\varepsilon$ as the binding energy of the pair of molecules 1 and 2 in the ensemble of N-2 other molecules.

## 4. Results and Discussion

### 4. A. Potential function and isotherm

The approach of our model, which relates the electronic properties to thermodynamic properties, is the asymptotes of the interaction potential at short and long ranges determined from the neutron scattering of alkali metals plus the corresponding theoretical calculations based on integral equation of state, in particular for liquid cesium.[10,11] Most theoretical investigations for thermodynamic properties of alkali metals are based on the low-density vapor, which dominantly consist of isolated atoms with bounded valance electrons (as well as small fractions of clusters with partially localized valence electrons). However, at high densities the valence electrons are dominantly delocalized and a giant molecule is encountered.[15] Treatment of the later is complicated but the nearly free electron model is applicable quite well. In this regard, the well-established pseudopotential concept,[16] which has been devised to study solids, is applied well to study the liquid alkali metals over wide ranges of temperatures and densities. A number of investigations have shown a high sensitivity of pair potential to the details of pseudopotential model.[17] At the same time many atomic properties such as phonon spectra and elastic constants, which are usually calculated to estimate the accuracy of model, are less sensitive to the pseudopotential model used.[17] Alkali metals are most popular in pseudopotential theory, and the method appropriately reproduces, for instant, experimental structure factors of the metals calculated in the Percus-Yevick (PY) approximation. For the heavy alkali metals (e.g. Rb and Cs) at the thermodynamic states $T < 0.7T_c$ the pseudopotential method still work well but marked deviations are noticed at $T > 0.7T_c$. These are not due to the failure of the PY approximation but because the pair potential becomes inappropriate at such low density.[17] At high temperature lager deviations from experimental structure factor is noticed in the case of heavier metals like liquid Cs. Although these methods are interesting by their first principle strength, their typical failure leave room for further studies of thermodynamic properties by the theories that uses semiempirical potential function.

The density dependence of liquid-metal potential function has been theoretically concerned.[7,8,11,12] Complications are raised when one is concerned with calculations of state properties of liquid metals by using molecular properties, because the pairwise additivity is simply applied rather accurately only to the respective low-density vapor for which the nature of molecular interaction remain constant as temperature is changed. The fact that the liquid alkali metals do not obey the law of corresponding states might be attributed to the divers changes in the nature of molecular interaction as their thermodynamic states are changed. Thus, the periodicity of the potential function at large r due to periodicity of lattice, regarded in the pseudopotential, is not valid at high temperature as it does at low temperatures. This observation is consistent with the potentials (at low and high T's) derived by inversion of structure factors in reference 11.

By the application of Eq. (3) [called $(6-3)$ potential function hereafter] and employment of thermodynamic equation of state, it is predicted that the isotherms $(Z-1)V^2$ for dense (fluid) cesium are linear functions of $(1/\rho)$. In Fig. 1, the isotherms of Eq. (9) [called $(6-3)$ isotherm hereafter] are plotted for liquid cesium in the range $350\,K - 2000\,K$ (the range $1100\,K - 2000\,K$ shown only). To construct the isotherms, PVT data has taken from Ref. 18. Isotherms are perfectly linear in the range $T_m - T_c$. For comparison, the same isotherms calculated using $(12-6)$ Lennard-Jones potential function [called $(12-6)$ isotherm hereafter] are shown in Fig. 2. The $(6-3)$ isotherm presented in this work applies quite well to liquid cesium over the whole liquid range, including metal-nonmetal transition and at $T_c$. The linear limit of an isotherm is estimated by $R^2 \geq 0.995$ limit, where R is the linear correlation coefficient. Comparison of Figs. 1 and 2 reveals that the interaction potential in liquid cesium, both at low temperature (dense liquid) and at high temperature (expanded liquid), is modeled accurately by $(6-3)$ potential so much the isotherms persist linear in both metallic and nonmetallic regions.

**4. B. Molecular parameters** Our semiempirical (potential) model based on structural data of experiment and investigation of the model will yield additional information about the structure of the system. Clearly this is valuable feature. The potential parameters $\varepsilon$ and $\sigma$ depend on both density and on the choice of potential model, and reasonably $(6-3)$ potential function introduces the correct asymptotes for interaction of cesium metal atom. With $m = 6$ for liquid cesium almost averages of repulsive asymptotes, determined from neutron scattering data,[11] are included in Eq. (3). This description of repulsive interaction is good enough to conclude the thermodynamic properties of liquid cesium because the high density limits, i.e., densities of compressed liquid close to the freezing temperature, correspond to the interatomic range located close to the equilibrium interatomic distance $r_m$. It worth noting that the available PVT data correspond to interatomic distances within the range $0.989 r_m < r < 1.406 r_m$. Highly compressed liquid densities data of cesium are needed to observe the application of $(6-3)$ isotherm in very small interatomic distances.

A smaller atomic diameter $\sigma$ is predicted by $(6-3)$ potential function than is predicted by the $(12-6)$ potential function. In this case ratio $\left[\sigma_{(6-3)}/\sigma_{(12-6)}\right] = 0.891$. Indeed an attraction proportional to $\left(1/r^3\right)$ at large r follows to account mainly for the attractions between ion-dipoles present in liquid cesium. However, we have not observed any effects due to absent of an oscillatory part in the potential function at long range, which has been concluded from the structure factor in the low temperature range.[10,11] It worth noting that the oscillation is the characteristic of the potential periodicity of the liquid lattice and is furnished by the pseudopotential model. In Fig. 3, $(6-3)$ potential function is compared with $(12-6)$ one. An appreciable softer repulsion and a more attraction of $(6-3)$ potential than $(12-6)$ potential is noticeable. We believe that the characteristic attraction of $(6-3)$ potential is mainly responsible for the persistent of the linear behavior of isotherms especially in the metal-nonmetal transition range and at critical temperature $T_c$.

The $(6-3)$ potential is a two parameters potential function, fairly analogue of Lennard-Jones (12-6) potential. It can be used to demonstrate law of corresponding states for selected thermodynamic properties, provided that the molecular parameters $\varepsilon$ and $\sigma$ are known and suitably follow $u^* = f(r^*)$ for a group of substances (alkali group here), where the reduced potential energy $u^* = u/\varepsilon$ and reduced interatomic separation $r^* = r/\sigma$. The success of law of corresponding states is guaranteed if the $\varepsilon$ and $\sigma$ are independent of the state of the system, a case which is not held perfectly specially for the condensed fluids. The variations of $\varepsilon$ and $\sigma$ with temperature for liquid metals and normal liquids have been already worked out.[7,8,11,12,19] In the present investigation the molecular parameters can be calculated using the numerical values of the linear parameters $B$ and $C$.

$$r_m = 2^{1/3}\left(\frac{-K^3}{2}\frac{C}{B}\right)^{1/3}, \qquad \varepsilon = \frac{RT}{N}\frac{B^2}{C}. \tag{11}$$

Note that the value of $C_2$ [see Eq. (10)] is practically small and thus is ignored. It is interesting to note that $\varepsilon$ is an explicit function of temperature and of linear parameters $B$ and $C$, which are expected to represent the interatomic interaction. On the other hand, in the temperature range $1950\,K < T < 350\,K$ the ratio $B^2/C$ for liquid cesium is within $3.4 - 38.8$ (i.e., one order of magnitude) indicating that a fair balance between interatomic attraction and repulsion is responsible for smooth variation of potential well-depth $\varepsilon$ with temperature [see Fig. (4)]. It is noteworthy to mention that $\varepsilon$ does not depend on the structure and the density as can be predicted by Eq. (11). However, $\sigma$ depends on both structure of the unit cell (e.g., K) and the interaction coefficients $B$ and $C$. As the temperature is increased, we do not know precisely how does the structure of liquid may differ from low temperature body-center-cubic structure, and thus we have kept K constant over the whole liquid range. Not very surprisingly, as the attraction coefficient $B$ is decreased $r_m$ is increased in accord to Eq. (11). Except a small wiggling around 1600 K, the variation of $r_m$ with temperature is linear; a plot of which is shown in Fig. 5. This observation is in accord with the neutron scattering data of liquid cesium form which the pair correlation function and thus the interatomic distance can be determined.[2]

**4. C. Comparison with experiments** The value of $r_m = 5.35\,\text{Å}$ at freezing temperature ($T_f = 303\,K$) which is close to the measurement,[20] $r_m = 5.40\,\text{Å}$, nonetheless it is increased by about 12% at the critical temperature. From spectroscopic measurements of the vapor state of cesium and *ab initio* calculations, the value of $\varepsilon/k$ is determined to be 1589.5 K, for which the contribution of both singlet and triplet type interactions are included.[21] In the range $350\,K - 2000\,K$, using the $(6-3)$ isotherm, $\varepsilon/k$ values are calculated, a plot of results is shown in Fig. 4. It is seen that $\varepsilon/k$ smoothly decreases with temperature and almost levels off from 1350 K on. This is a characteristic temperature, which we will refer to latter.

Now according to our model, taking into account only the interaction of a central cesium atom with 8 nearest adjacent atoms, the value of $\varepsilon/k$ at the freezing temperature is estimated to be equal $1710.5 K$, which is higher by about 7.6% with respect to spectroscopic data.[21] The calculation of $\varepsilon$ by our method depends much on the potential model used to establish the relation for which the experimental liquid PVT data have been the input data. This method for estimation of molecular parameters is simple and less costly than other methods, which somewhat depend of complicated experimental set up.

The $(6-3)$ potential is an effective potential that includes not only a pair interaction but also many-body interactions and therefore it is not always comparable with a pair potential.

**4. D. Analogue potential functions** The metal-nonmetal transition is particularly of interest in this work. To show the range of linear behavior of Eq. (9), $R^2$'s are plotted versus temperature in Fig. 6. It is seen that over the whole liquid range $R^2 \geq 0.995$ with some wiggling in the transition range. Additionally, to see the effects of slopes of attraction and repulsion of a potential function on the linear behavior of the isotherms, $R^2$'s are also plotted as a function n in Fig. 6, and as a function m in Fig. 7. For comparison, in both figures $R^2$'s of $(6-3)$ isotherms and $(12-6)$ isotherms are included. For $(6-3)$ isotherms $R^2$'s are determined using Eq. (9). For other combination of m and n, $R^2$'s are determined by using Eq. (1) to derive an isotherm of the general form

$$(Z-1)V^{n/3} = c + b\rho^{(m-n)/3}. \tag{12}$$

According to choices of m and n, Eq. (12) is used to produce Figs. 6 and 7 for liquid cesium. Notice that for m=6 and n=3, Eqs. (9) and (12) are identical except for the sign of the slope and do not have effects on the conclusions we are seeking for. Examination of Figs. 6 and 7 reveals that, with $(6-3)$ potential function, linear isotherms persist over the whole liquid range including metal-nonmetal transition range, which onsets at the temperature 1350 K, and at the $T_c (=1924 K)$.[1] Any potential function harder than $(6-3)$ potential will produce an isotherm that deviates from the linear behavior more than $(6-3)$ isotherm over the whole liquid range. In general, in the transition range and at the critical temperature larger deviations are seen. In all case the deviations from linear behavior onset at a higher temperature than that of $(12-6)$ isotherm (for which onsets at 1350 K). Thus a simple conclusion is that the harder the potential function the larger the deviations from linear behavior in the transition region. The same is true for any potential that includes less attraction than $(6-3)$ potential function.

One of the most striking results of the measurement of liquid-vapor coexistence curve of Rb and Cs,[22] was the demonstration of marked departure from law of rectilinear diameter.[23,24] It has been proposed that the reason resides in the polarization term predicted by Blazej and March in the effective pair potential u(r).[25] It leads to an attractive (oscillatory) asymptotic contribution to u(r) proportional to $(1/r^4)$ at sufficiently large interatomic distance. This is in contrast to the form $(\cos(2k_f r)/r^3)$, where $k_f$ is the Fermi wave number of conduction electrons. The above proposal has been justified as follows: the mean free path of electron is finite, which from the Heisenberg uncertainty principle leads to blurring of the Fermi surface by an amounts $\Delta k_f$, a factor proportional to $e^{-\Delta k_f r}$ will eventually cause a dumping of the oscillatory part $(\cos(2k_f r)/r^3)$ at sufficiently large r and polarization contribution proportional to $(1/r^4)$ will dominate.[26] Our procedure presented in this work, however, shows that, (*i*) not including oscillation in the potential function at long range does not introduce detectable effects on the linearity of the (6-3) isotherm in expanded liquid cesium, (*ii*) any potential function which introduce less attraction than $(1/r^3)$ at long range, results in an isotherm which deviates from linear behavior in the expanded liquid cesium. Hence, for seek of theoretical investigations, the dumping factor $e^{-\Delta k_f r}$ becomes invalid at high T (see figures 6 and 7).

**4. E. Role of linear parameters** The slope of the $(6-3)$ isotherm B is related to the attractive part of the $(6-3)$ potential function and thus, B is expected to conform to the second virial coefficient $B_2$ as the intercept A of the original LIR does in the case of normal fluids.[5,7,8] The plot of B versus temperature for liquid cesium is shown in Fig. 8. No reliable experimental $B_2$ values are available for comparison. However, we have previously reported $B_2$'s of alkali metals[27] calculated by using the available diatom fractions of the metal vapor at low-pressure,[18] for which a simple relation is manipulated between equilibrium constant of the diatom formation process and $B_2$. These data plus the $B_2$'s calculated by integration using pair potential functions for singlet and triplet of atomic-collision states, corrected for the quantum effects up to the first order quantum correction,[28] are also shown in Fig. 8. By the plots in Fig. 8 we meant to imply an interesting observation that at high temperatures the slope B is in the same order of magnitude as $B_2$'s calculated by the two methods mentioned above. At the critical point, where the average coordination number in liquid cesium is known to be almost 2, they compare well. The recent molecular dynamic studies of liquid cesium for the pair potential has came up to a similar conclusion.[29] The result of this studies have been used to calculate the bulk modulus, which the reciprocal isothermal compressibility. As the temperature increases, the theoretical (computer simulation) bulk modulus becomes more agreeable with the experimental one, and at $T_c$ the differences become less that the experimental error. This has been attributed to the contribution of the electron gas which diminishes as the metallic character is replace by non-metallic one in the expanded liquid metal. Thus, as $T_c$ is approached electron gas make no appreciable contribution to compressibility indicating localization of valence electrons and formation of small molecular clusters.

The property of the slope B can be studied further by rearrangement of the $(6-3)$ isotherm, which leads to the following virial (like) equation of state:

$$Z = 1 + B\rho + C\rho^2. \tag{13}$$

In Eqs. (9) and (13), terms B and C are used to imply that second and third virial (like) coefficients, respectively, may be traced from attraction and repulsion of the $(6-3)$ potential function. On the contrary (12-6) isotherm (e.g., the original LIR) on rearrangement will produce a virial equation of state up to fifth virial coefficient with the second and forth virial coefficients missing. To the authors knowledge's, this is the characteristic that is observed (*i*) using the procedure followed plus $(6-3)$ potential function only, and (*ii*) is obtained regardless of fluid under investigation, recalling however this particularly might be applicable only to those fluid systems obeying the model presented in this work. The further generalizations of these results are under investigation.

The Eq. (13) can be analyzed numerically for the contribution of attraction and repulsion interactions. Contributions of these two interaction terms are shown in Table II. Here two points are noticeable. The first is that, in all cases the attraction term and the repulsion term is not only in the same order of magnitude but also very close to each other. The second, if for the time being we treat the equation (13) as a truncated series, it is seen that at low temperature the series is likely to diverge, while likely it converges at high temperatures. Either convergence or divergence is slow. The second feature above is nicely predicted by theory and is attributed to the characteristics of the (6-3) potential. The potential function with the term $(1/r^3)$ cause that the integral

$$\int r^2 g(r,T)(\partial u/\partial r)dr$$

changes slowly with r. The integral diverges slightly only at low T where $g(r,T)$ corresponds to a highly structured fluid, and converges slowly at high temperature where $g(r,T)$ has the characteristics of an expanding fluid. These are consistent with the numerical data in Table II. Hence prediction of accurate thermodynamic properties seems to be logical because the Eq. (13) gives the whole contributions (without a reminder, which is the specification of a truncated (real) virial series).

The perfect linearity of the isotherm in the whole liquid range permits extrapolation of the equation of state to the region of density where the experimental difficulties do not allow measurements. This plus the fact that the slope B and the intercept C are linear functions of 1/T [see Figs. 9(a) and 9(b)] indicate the advantages of the isotherm derived using $(6-3)$ potential function.

## 5. Conclusions

In this work, the two-parameter $(6-3)$ potential function has introduced for compressed liquid cesium and according to that, the exact thermodynamic equation of state has been solved. $(6-3)$ potential function accounts appropriately for the large attraction at long range and a soft repulsion at the short range. The equation of state, introduced as linear isotherm, persist the linear behavior in dense region, where typically liquid cesium has a solid like matrix structure, and in the low density (expanded liquid), where it changes gradually from metal character into a non-metal kind. It persists linear even at the critical temperature, where rarely an equation of state acts analytically. The attraction of $(6-3)$ potential is overwhelmingly large such that it seems to be a (lump) sum for ion-ion interaction in high density liquid plus all sort of long range multipolar interactions present in the expanded liquid cesium. By this approach, the potential periodicity of the lattice at large r can be included, though it leaves no detectable effect on the calculated thermodynamic properties. The molecular parameters of the potential function $r_m$ and $\varepsilon$ have calculated from the PVT data of liquid state. The value of $r_m$ depends explicitly on the linear parameters $B, C$ (e.g., interaction parameters), and the structure of unit cell K. It increases quite steadily with temperature. The model of this work has treated $\varepsilon$ as the binding energy of the pair of atoms 1 and 2 in the ensemble of N-2 other atoms of the liquid and accordingly found as an explicit function of $B, C$, and temperature. At the freezing temperature of cesium, the $(6-3)$ isotherm reproduces $r_m$ within $0.9\%$ of the experimental value.

The $(6-3)$ potential function is a reliably applicable in the range of freezing to critical temperature. The existence of a definite temperature dependence of (6-3) potential function, which is due to an appreciable decrease in the density with increasing temperature, could serve as an additional criterion for the model constructed. This work is also a check of our pervious work in which exp-6 potential has applied for the purpose of accounting suitable attraction and repulsion asymptotes.[8] It is more favorable than exp-6 isotherm because, (*i*) it is more accurate in the transition region and at $T_c$, (*ii*) exp-6 isotherm depends explicitly on $r_m$ that is not a constant as the temperature is changed, and (*iii*) the results of this work substantiates further quantitative investigations, which are undertaken, in order to conclude its physical insight and its impact on the thermodynamics of liquid alkali metals.

**Acknowledgement**

Supports by the Research Committee of Shiraz University are greatly acknowledged. Thanks are due to Dr. E. Keshavarzi for providing preprint of reference 7 and to Professor F.E. Lays for providing the preprint of reference 26.

Table 1. The linear parameters of the isotherm 9 for liquid cesium.

| T | $B \times 10^3$ $(m^3, mol^{-1})$ | $C \times 10^7$ $(m^6, mol^{-2})$ | $R^2$ | $\frac{\Delta P}{bar}$ | $\frac{\Delta \rho}{(gr, cm^{-3})}$ |
|---|---|---|---|---|---|
| 350 | -2.7785 | 1.9885 | 0.9992 | 50-600 | 1.815-1.880 |
| 400 | -2.3961 | 1.7332 | 0.9998 | " | 1.787-1.854 |
| 450 | -2.1228 | 1.5524 | 0.9999 | " | 1.759-1.825 |
| 500 | -1.8552 | 1.3711 | 0.9997 | " | 1.730-1.803 |
| 550 | -1.6758 | 1.2526 | 0.9997 | " | 1.702-1.778 |
| 600 | -1.4864 | 1.1229 | 0.9995 | " | 1.673-1.753 |
| 650 | -1.3367 | 1.0199 | 0.9998 | " | 1.645-1.728 |
| 700 | -1.2324 | 0.9503 | 0.9998 | " | 1.617-1.703 |
| 750 | -1.1496 | 0.8962 | 0.9999 | " | 1.589-1.675 |
| 800 | -1.0756 | 0.8480 | 0.9998 | " | 1.561-1.652 |
| 850 | -0.9949 | 0.7922 | 0.9998 | " | 1.533-1.623 |
| 900 | -0.9288 | 0.7474 | 0.9997 | " | 1.505-1.603 |
| 950 | -0.8703 | 0.7076 | 0.9998 | " | 1.476-1.578 |
| 1000 | -0.8283 | 0.6812 | 0.9999 | " | 1.447-1.552 |
| 1050 | -0.7715 | 0.6411 | 0.9999 | " | 1.416-1.527 |
| 1100 | -0.7205 | 0.6043 | 1.0000 | " | 1.385-1.503 |
| 1150 | -0.6748 | 0.5716 | 1.0000 | " | 1.352-1.478 |
| 1200 | -0.6359 | 0.5440 | 0.9999 | " | 1.320-1.453 |
| 1250 | -0.5967 | 0.5152 | 0.9999 | " | 1.287-1.429 |
| 1300 | -0.5627 | 0.4905 | 0.9996 | " | 1.254-1.404 |
| 1350 | -0.5319 | 0.4677 | 0.9994 | " | 1.220-1.380 |
| 1400 | -0.5080 | 0.4513 | 0.9993 | " | 1.186-1.355 |
| 1450 | -0.4882 | 0.4383 | 0.9992 | " | 1.151-1.330 |
| 1500 | -0.4674 | 0.4236 | 0.9993 | " | 1.115-1.305 |
| 1550 | -0.4515 | 0.4138 | 0.9992 | " | 1.078-1.280 |
| 1600 | -0.4351 | 0.4029 | 0.9990 | " | 1.039-1.255 |
| 1650 | -0.4161 | 0.3885 | 0.9992 | " | 0.994-1.230 |
| 1700 | -0.3923 | 0.3678 | 0.9997 | 100-600 | 0.976-1.205 |
| 1750 | -0.3797 | 0.3597 | 0.9998 | " | 0.926-1.179 |

| 1800 | -0.3692 | 0.3539 | 0.9999 | " | 0.873-1.154 |
| 1850 | -0.3584 | 0.3474 | 1.0000 | " | 0.811-1.127 |
| 1900 | -0.3504 | 0.3438 | 1.0000 | " | 0.742-1.101 |
| 1950 | -0.3475 | 0.3474 | 0.9997 | " | 0.655-1.072 |
| 2000 | -0.3002 | 0.2294 | 0.9982 | 200-600 | 0.792-1.045 |

Table 2. Numerical analysis of Eq. (13)

| T (K) | P Bar | $Z = 1 + B\rho + C\rho^2$ |
|---|---|---|
| 400 | 100 | $1 - 32.34 + 31.58$ |
|  | 300 | $1 - 32.78 + 32.42$ |
|  | 600 | $1 - 33.42 + 33.72$ |
| 1000 | 100 | $1 - 9.079 + 8.186$ |
|  | 300 | $1 - 9.329 + 8.642$ |
|  | 600 | $1 - 9.683 + 9.312$ |
| 1900 | 100 | $1 - 1.853 + 0.919$ |
|  | 300 | $1 - 2.515 + 1.790$ |
|  | 600 | $1 - 2.900 + 2.380$ |

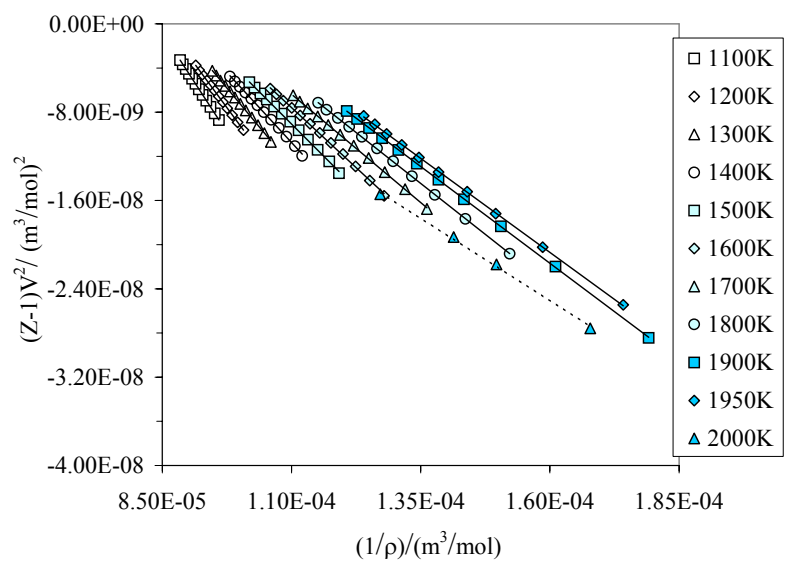

**Figure 1**. The (6-3) isotherms for liquid cesium in the range of freezing to critical temperature (the range $1100\,\text{K} - 2000\,\text{K}$ is shown only).

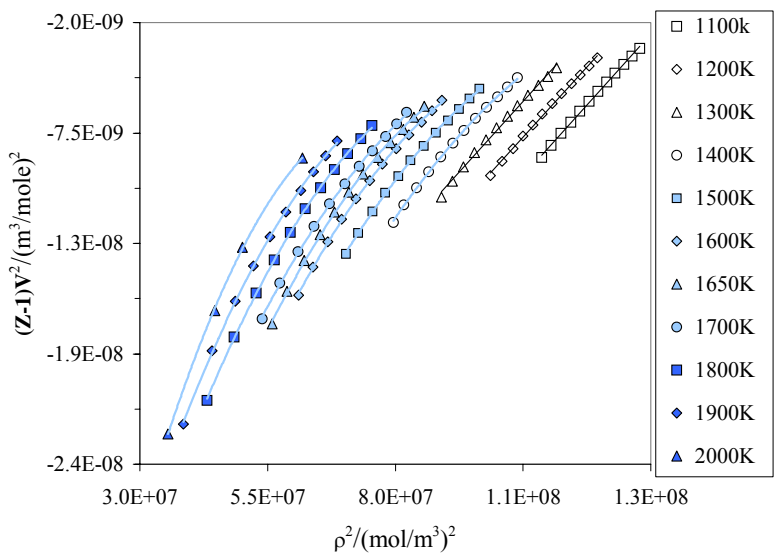

**Figure 2**. The (12-6) isotherms for liquid cesium from freezing to critical temperature (the range $1100\,\text{K} - 2000\,\text{K}$ is shown only). Gray lines are used to show the range of deviation from linear behavior.

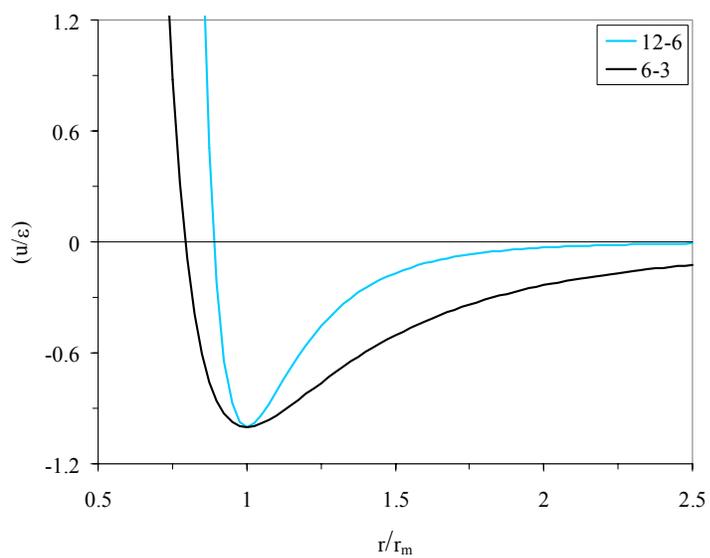

**Figure 3**. Plots of (6-3) and (12-6) potential functions

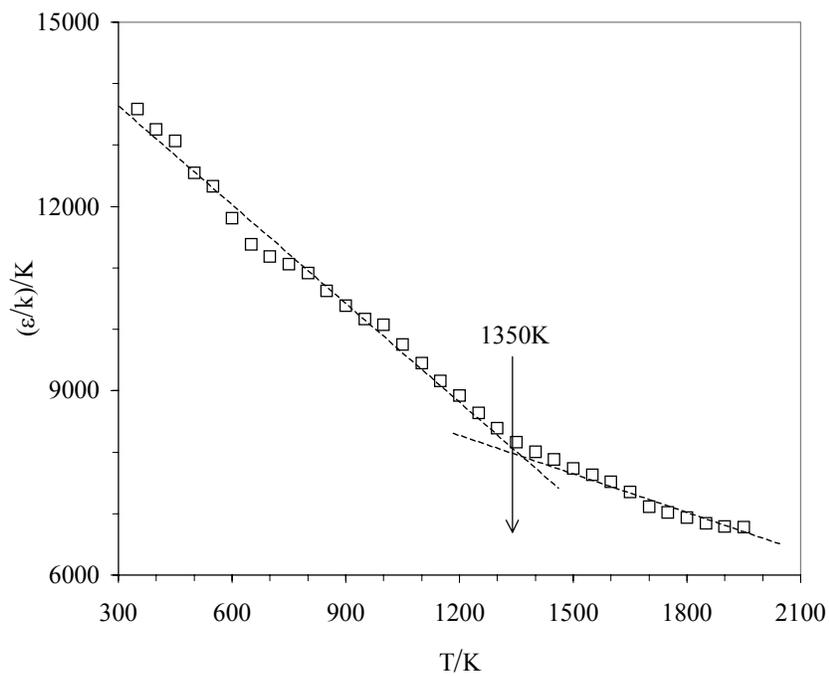

**Figure 4.** Plot of ε/k versus temperature for liquid cesium. The dashed line has drawn through to emphases the trends around the onset of transition region.

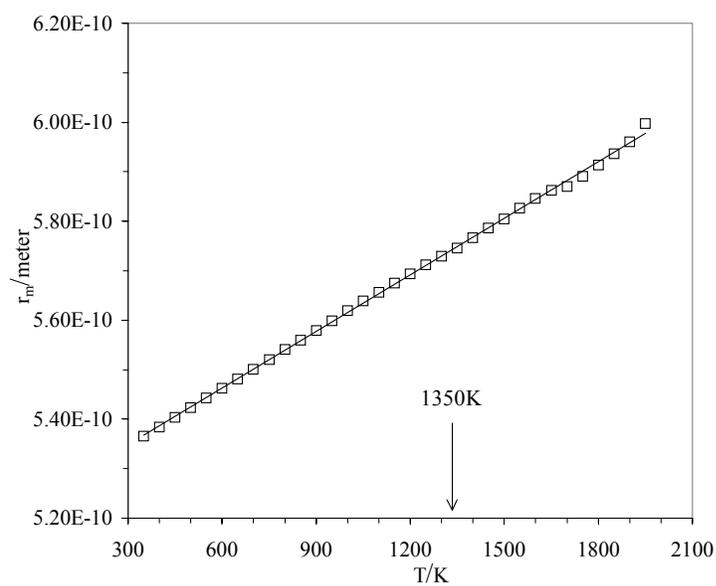

**Figure 5.** Plot of $r_m$ versus temperature for liquid cesium. The solid line is the linear fit to $r_m$ values.

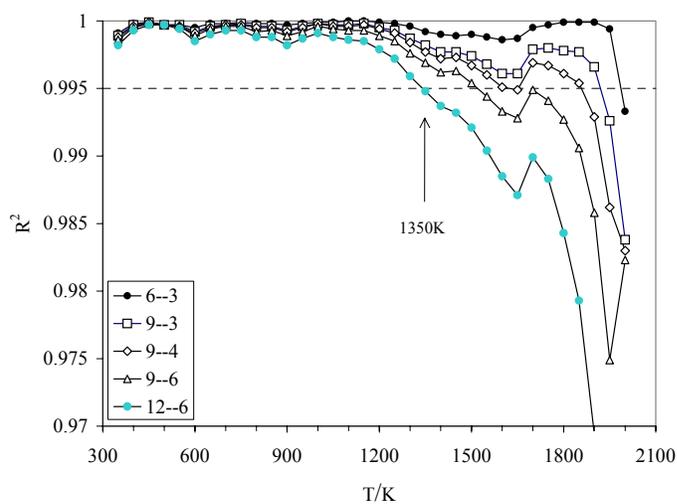

**Figure 6.** Plots of $R^2$ of (6-3) isotherms for liquid cesium in the temperature range $350\,K - 2000\,K$. The related $R^2$ for potential functions with $m = 9$ and $n = 3,4,6$ have examined (see the text). Plots for (12-6) isotherm are included for comparison.

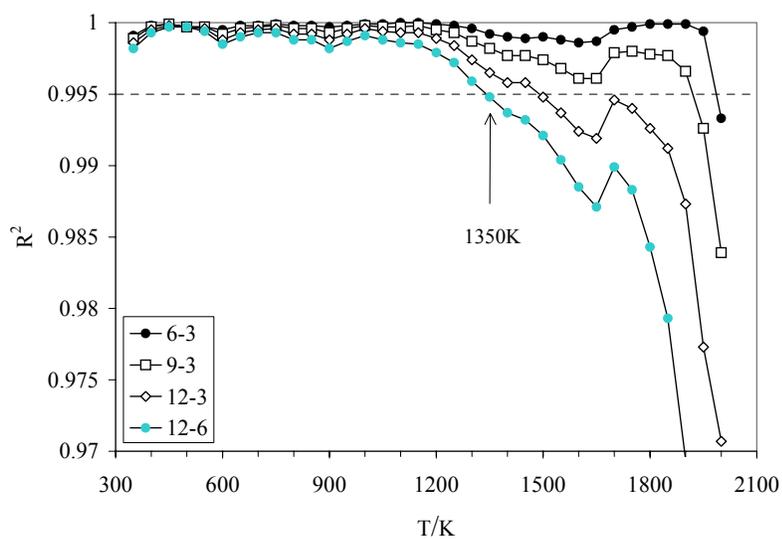

**Figure 7**. The same as the figure 6 except for $n = 3$ and $m = 9, 12$ (see the text).

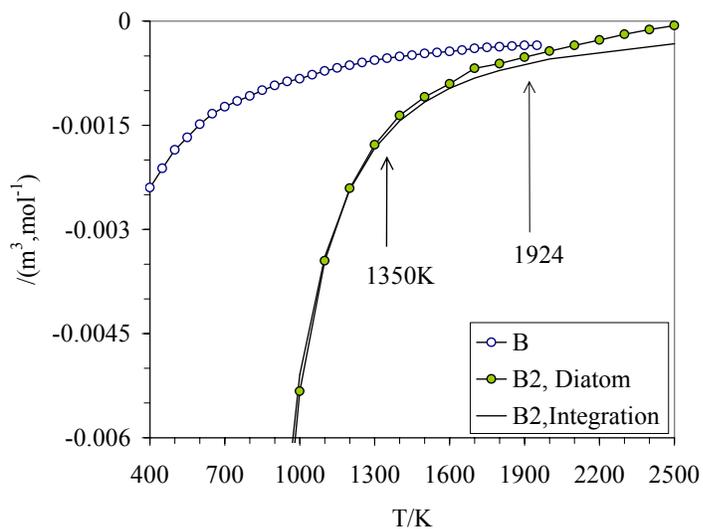

**Figure 8**. Plot of slope B versus temperature for liquid cesium (see text for other two plots).

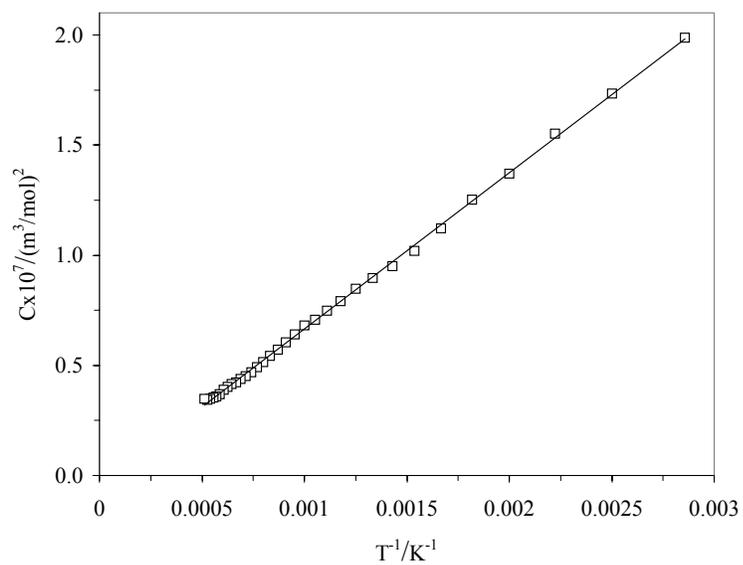

**Figure 9a.** Plot of B versus inverse temperature. Solid line is the linear fit to the B data points (squares).

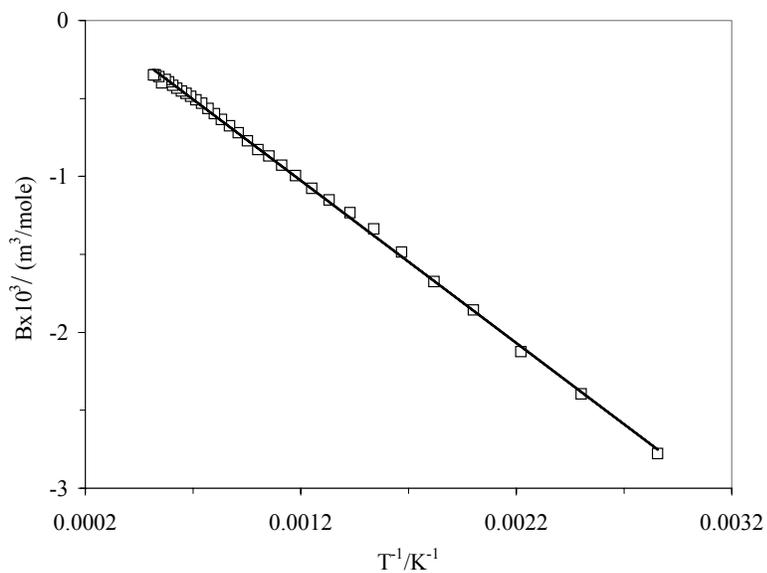

**Figure 9b.** Plot of C versus inverse temperature. Solid line is the linear fit.

24